# Silicon Nitride Stress Liner Impacts on the Electrical Characteristics of AlGaN/GaN HEMTs

Wei-Chih Cheng,[1,2] Tao Fang,[1] Siqi Lei,[1] Yunlong Zhao,[3] Minghao He,[1] Mansun Chan,[2]
Guangrui (Maggie) Xia,[3] Feng Zhao,[4] and Hongyu Yu[1,5,6]

[1] SUSTech School of Microelectronics, Southern University of Science and Technology, Shenzhen, China
[2] Department of Electronic and Computer Engineering, Hong Kong University of Science and Technology, Hong Kong, China
[3] Department of Materials Engineering, The University of British Columbia, Vancouver, BC, Canada
[4] School of Engineering and Computer Science, Washington State University, Vancouver, WA, USA
[5] Shenzhen Key Laboratory of the Third Generation Semi-conductor, Shenzhen, Guangdong, China
[6] GaN Device Engineering Technology Research Center of Guangdong, Shenzhen, Guangdong, China
E-mail: yuhy@sustc.edu.cn

*Abstract*—Due to the piezoelectric nature of GaN, the 2DEG in AlGaN/GaN HEMT could be engineered by strain. In this work, SiN$_x$ deposited using dual-frequency PECVD was used as a stressor. The output performance of the devices was dominated by the surface passivation instead of the stress effect. However, the threshold voltage was increased by the induced stress, supporting strain engineering as an effective approach to pursue the normally-off operation of AlGaN/GaN HEMTs.

## I. Introduction (Heading 1)

GaN-based high electron mobility transistors (HEMTs) have emerged as one of the key device options for next generation power electronics and high frequency applications. The wide bandgap provides the ability to support high electric fields. Meanwhile, a high on-current ($I_{on}$) can be achieved due to the high saturation speed of electrons in GaN and the presence of two-dimensional electron gas (2DEG) at the AlGaN/GaN heterojunction.

One of the main reasons of forming 2DEG at the AlGaN/GaN heterojunction is the piezoelectric property of GaN. The lattice-mismatch stress between AlGaN and GaN can generate a net internal electric field in the AlGaN barrier, which drives electrons to accumulate in the quantum well near the heterojunction [1]. Due to the piezoelectric nature of GaN, the 2DEG concentration and thus the electrical characteristics of the AlGaN/GaN HEMT can be modulated by applying strain [2-3]. Besides, simulation results showed that the strain provided by the passivation layer can induce significant amount of piezoelectric charges in the submicron gate region [3]. Therefore, strain engineering is believed to be an effective approach to adjust the threshold voltage ($V_{th}$) of an AlGaN/GaN HEMT with a scaled gate length.

In our previous work, AlGaN/GaN HEMTs with micron gate lengths (micron devices) were prepared, and SiN$_x$ stressors with different intrinsic stresses were deposited on the devices by dual-frequency plasma-enhanced chemical vapor deposition (PECVD) [4]. Additional stress in AlGaN layer was observed using Raman spectroscopy [4]. However, surface damages caused by the deposition process was believed to dominate the output performance change. The results will be reviewed in sec. II. Besides, devices with sub-micron gate lengths (submicron devices) and different strain SiN$_x$ were further investigated. Also, we used compressive strain in the 2DEG region underneath the gate to achieve a $V_{th}$ increase of approximately one volt in a device with a L$_g$ = 100 nm, giving an alternative way to pursue normally-off devices for RF applications.

## II. Impacts from SiNx Stressors on Micron AlGaN/GaN HEMTs

The structure of the AlGaN/GaN HEMT in this work is shown in Fig. 1. The strain was provided by the SiN$_x$ passivation layer deposited by PECVD with dual plasma excitation frequencies. By varying the duty cycles of the high and low plasma excitation frequencies throughout the deposition process, the strain of the deposited SiN$_x$ could be engineered [5]. Three sets of frequency duty cycle parameters were used in this work, corresponding to tensile, compressive, and unstrained SiN$_x$, respectively. The setting details were explained in Ref. [4].

The on-resistance ($R_{on}$) of the devices with different SiN$_x$ stresses are summarized in Table I with the 2DEG sheet resistance extracted by the transmission line model ($R_{\square,2DEG}$). The tensile SiN$_x$ was expected to introduce an additional compression to AlGaN and lead to the increased $R_{on}$ and $R_{\square,2DEG}$. However, the opposite behavior was observed. To investigate this problem, Raman spectroscopy was used to measure the stress in the AlGaN and GaN layers.

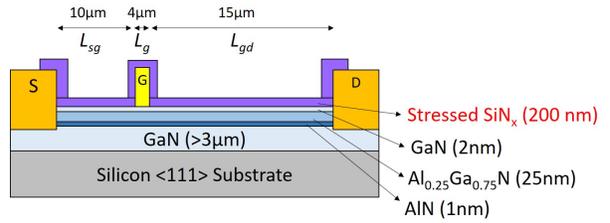

Fig. 1. The device structure of AlGaN/GaN HEMT with the gate length ($L_g$), source-to-gate length ($L_{sg}$,) and gate-to-drain length ($L_{gd}$) labelled.

TABLE I. Channel Resistances of Micron Devices

| SiN$_x$ | $R_{\square,2DEG}$ ($\Omega$) | $R_{on}$ ($\Omega$-mm) |
|---|---|---|
| Compressive | 1082 | 32 |
| Unstrained | 356 | 12 |
| Tensile | 299 | 11 |

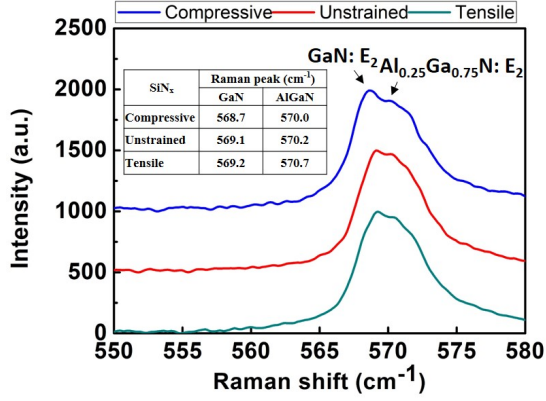

Fig. 2. Raman spectra of AlGaN/GaN heterostructure with different strains of $SiN_x$. The AlGaN peak shifted toward a higher frequency due to the compressive strain induced in AlGaN by the tensile $SiN_x$ on the surface.

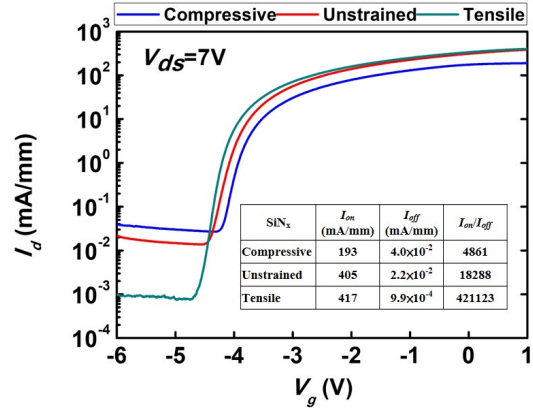

Fig. 3. Transfer characteristics of AlGaN/GaN HEMTs with different strain $SiN_x$. The device with compression $SiN_x$ showed a larger $I_{off}$ and a lower $I_{on}$. The $I_{on}$, $I_{off}$ are defined as the $I_d$ at $V_g = 1$ and -6 V, respectively

The $E_2$ phonon frequency decreases with tensile stresses [6]. In this study, the access regions of devices (between gate and drain) with different stressed $SiN_x$ were probed with a 442-nm laser, and each Raman spectrum showed a GaN peak (near 569 cm$^{-1}$) and an AlGaN peak (near 570.5 cm$^{-1}$), as shown in Fig. 2. The AlGaN peak of the sample with tensile $SiN_x$ appeared at a higher frequency than the other samples, suggesting that additional compression was generated in the AlGaN layer. That translates to an increased $R_{on}$ and $R_{\square,2DEG}$. Therefore, the experimental results showing a decreased $R_{on}$ and $R_{\square,2DEG}$ could not be explained by the stress effect.

Another reason to explain the experimental observations is the surface passivation quality. The compressive PECVD $SiN_x$ was prepared using longer period of low-frequency plasma excitation, which may cause more damages to the AlGaN surface by nitrogen ion bombardment [7]. As a result, the higher off current ($I_{off}$) and lower $I_{on}$ were observed in the device with compressive $SiN_x$, as shown in Fig. 3. In contrast, the tensile $SiN_x$ was deposited using high-frequency plasma excitation throughout the deposition, which should have less ion bombardment and a better passivation quality to AlGaN. Therefore, the device with tensile $SiN_x$ can reach the lower $R_{on}$ and $R_{\square,2DEG}$, even though the additional compression was introduced to AlGaN.

In summary, the strain and the surface quality effects from the $SiN_x$ stressors are two competing effects. In this case, the output performance (especially $I_{off}$) of devices were dominated by the passivation quality instead of the stress of $SiN_x$. Besides output characteristics, a positive 0.3-volt $V_{th}$ shift was observed when a compressive $SiN_x$ was applied on the devices with $L_g = 4$ μm. In the next section, the strain effect on the $V_{th}$ of the device with different $L_g$ (0.1 to 4 μm) will be explored

### III. $V_{TH}$ INCREASE IN SUBMICRON ALGAN/GAN HEMTS FOR RF APPLICATION USING STRAIN ENGINEERING

The device structure in the study is similar to that shown in Fig. 1, with $L_g$ from 0.1 to 4 μm, as shown in Table II. The submicron devices exhibited a reduced $R_{on}$ upon application of the tensile $SiN_x$, consistent with the behavior observed in the micron devices.

The gate-length dependence of $V_{th}$ voltage of devices with different stressed $SiN_x$ was drawn in Fig. 4, including the micron devices in Sec. II. Devices with unstrained $SiN_x$ exhibited a lower $V_{th}$ when gate length was scaled more. The more negative $V_{th}$ of unstrained devices was due to short-channel effects or gate-fringing effect [8]. In contrast, the devices with compressive $SiN_x$ exhibited a $V_{th}$ increase when the gate length was scaled.

To investigate the gate length dependence of the induced stress underneath the gate region, the stress distributions were simulated using Silvaco. In the simulations, $SiN_x$ layers with -1 GPa intrinsic stress were applied to the devices with $L_g = 0.1$ and 0.4 μm, as shown in Fig. 5. For devices with $L_g = 0.1$ μm, more intense compression is produced at the edge of the gate. In addition, the compressive $SiN_x$ has a greater influence over the heterostructure under the gate region, while tensile stress exists in some regions of under the 0.4 μm gate. Because the smaller gate can be influenced more by the stressed $SiN_x$, the $V_{th}$ of devices increased when gate length was decreased.

TABLE II. DEVICE DIMENSIONS

| Device type | $L_g$ (μm) | $L_{sg}$ (μm) | $L_{gd}$ (μm) |
|---|---|---|---|
| Submicron device | 0.1 | 3.7 | 6.3 |
|  | 0.2 | 3.7 | 6.3 |
|  | 0.4 | 3.7 | 6.3 |
|  | 0.4 | 3.7 | 10.3 |
| Micron device | 4 | 10 | 15 |

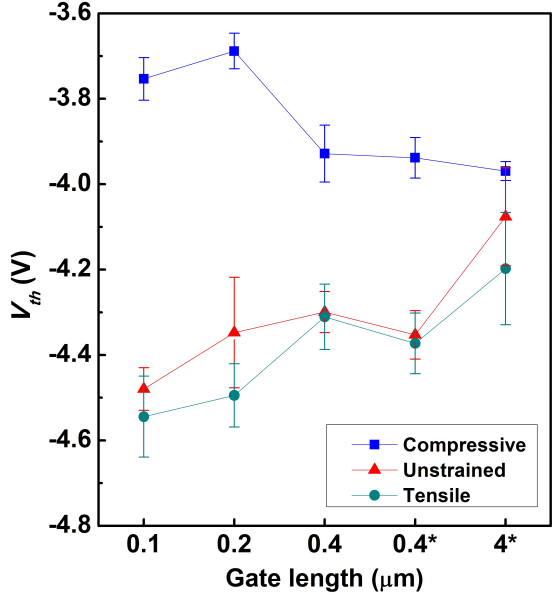

Fig. 4. Gate-length dependence of the $V_{th}$ of the devices with different stressed $SiN_x$. The error bars are also showed, representing the standard deviation out of the data point.

According to the stress simulation results (Fig. 5), induced stress underneath the gate is larger for smaller $L_g$. Compressive stressors introduce compressive in-plane stresses in AlGaN underneath the gate, which causes a smaller amount of piezoelectric charges in AlGaN, depleting the 2DEG under the metal gate. As a result, even if the short channel effect had an influence, a submicron device with a compressive $SiN_x$ liner could achieve a higher $V_{th}$.

In summary, for these devices with different $L_g$, the $V_{th}$ dependence can come from two sources: the short-channel effect and the stress effect. The experimental data support that the stress effect is the dominant effect.

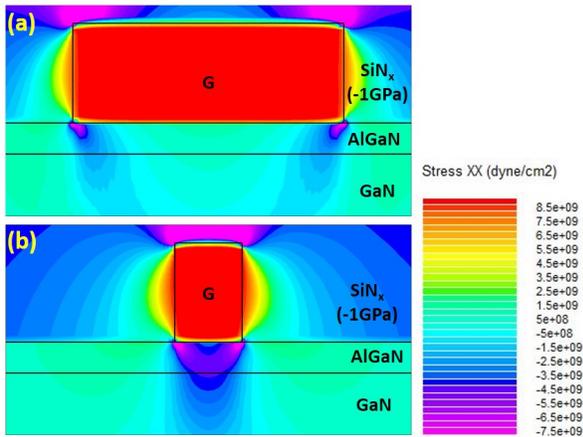

Fig. 5. Simulated stress distributions in the devices with 1-GPa compressive $SiN_x$ liner. (a) $L_g = 0.1$ μm and (b) $L_g = 0.4$ μm. In the device with 0.1 μm gate length, more intensive compression is developed at the gate edge. There are more compressive stresses in the AlGaN/GaN region underneath the gate.

For RF application, normally-off operation is preferred. An etching process, which is necessary to prepare enhancement-mode device, such as MIS-HEMT and P-GaN HEMT, is undesired for high frequency application [9]. The fact that the $V_{th}$ of scaled devices can be adjusted by strain engineering without an etch step provides a useful approach for pursing enhancement-mode AlGaN/GaN HEMT.

## IV. CONCLUSION

Due to the piezoelectric nature of GaN, the 2DEG in AlGaN/GaN HEMT could be engineered by stressed $SiN_x$. The output performance ($I_{on}$ and $R_{on}$) of the devices was dominated by the surface passivation instead of the stress effect. However, the $V_{th}$ could be increased by the induced stress. It was increased by approximately 1V with a compressive $SiN_x$ liner for devices with $L_g = 0.1$ μm. This finding supports that strain engineering is an effective approach pursuing enhancement-mode AlGaN/GaN HEMTs.

ACKNOWLEDGMENT

This work was supported by the projects with grant number 2017A050506002, JCYJ20160226192639004, and JCYJ20170412153356899. Device fabrication was performed at Materials Characterization and Preparation Center, Southern University of Science and Technology with equipment supported with various grants.